\documentclass[twocolumn,showpacs,preprintnumbers,amsmath,amssymb,aps]{revtex4}
\usepackage{graphicx}
\usepackage{dcolumn}
\usepackage{bm}
\begin{document}
\title{Exotic fission properties of highly neutron-rich Uranium isotopes}
\author{L. Satpathy, S.K. Patra and R.K. Choudhury$^1$}
\affiliation { Institute of Physics, Sachivalaya Marg, Bhubaneswar-751 005, 
India}
\affiliation { $^1$Bhabha Atomic Research Centre, Nuclear Physics Division,
Mumbai-400 085, India}
\begin{abstract}

The series of Uranium isotopes with $N=154 \sim 172$ around the magic number
N=162/164 are identified to be thermally fissile. The thermal neutron fission
of a typical representative $^{249}$U of this region amenable to synthesis in
the radioactive ion beam facilities is considered here.  
Semiempirical study of fission barrier height
and width shows this nucleus to be infinitely stable against
spontaneous fission due to increase in barrier width arising out of excess
neutrons. 
Calculation of probability of fragment
mass yields and microscopic study in relativistic mean field theory,
show this nucleus to undergo a new mode of thermal fission decay termed 
{\it multifragmentation fission} where a number of prompt scission
neutrons are simultaneously released along with the two
heavy fission fragments.
\end{abstract}
\maketitle

Exploration of nuclear landscape with radioactive ion beam facilities
is currently underway in several laboratories around the world.
Presently about 2000 nuclei are known, and different mass models
predict the survival of another 5000 nuclei with
varying life time which could be synthesised in the laboratory \cite{thoen04}.
Most of the nuclei will occur in the neutron-rich side of the nuclear
chart. Many of these nuclei will have unusual neutron to proton ratio,
quite different from the usual ones in the valley of stability.
Therefore with the synthesis of such nuclei one may expect
to observe new nuclear phenomena and novel features of nuclear
dynamics.  The two Uranium isotopes $^{233}$U and $^{235}$U 
and Plutonium isotope
$^{239}$Pu in the actinide region are suitable for energy production being
thermally fissile in nature. Here we investigate to see if
heavier neutron-rich isotopes of Uranium could exist having
thermally fissile properties, and if so what would be their fission
decay properties. 

On the basis of fission barrier $B_f$ and neutron separation energy
$S_n$ systematics, we find that the chain of Uranium isotopes with
neutron number N=154 $\sim$172 possess the thermally fissile property
\cite{howard80}. These isotopes span on either side of N=162/164
which has been predicted to be magic in numerous theoretical studies carried
over the years \cite{cwiok96}. The effect of close shell manifests in
relatively higher values of $S_n$ and lower values of $B_f$ (favouring more
bound compact configuration) rendering the system thermally
fissile, which may be the case with the above series of Uranium
isotopes. All these nuclei are stable against
alpha decay and several of them in the lower side have beta-decay half-life
of several tens of seconds \cite{moller97}. 
\begin{table}
\centering
\caption{Fission barrier $B_f$ (in MeV) obtained in finite-range droplet model
(FRDM) \cite{moller95} and Howard and M\"oller (H.M.) \cite{howard80}
are compared with experimental data \cite{moller95}.
}
\label{table1}
\begin{tabular}{ccccccc}
\hline
\multicolumn{1}{c}{nuclei} & \multicolumn{1}{c}{$B_f$ (expt.)}
& \multicolumn{1}{c}{$B_f$ (H.M.)}
& \multicolumn{1}{c}{$B_f$ (FRDM)}  \\
 \hline
 $^{228}$Th &6.50&  &7.43 &   \\
 $^{230}$Th &7.00& 7.79  &7.57   \\
 $^{232}$Th &6.30&7.40  &7.63   \\
 $^{234}$Th &6.65& 6.83  &7.44   \\
 $^{232}$U &5.40& 6.31  &6.61   \\
 $^{234}$U &5.80& 6.03  &6.79   \\
 $^{236}$U &5.75& 5.74  &6.65   \\
 $^{238}$U &5.90& 5.83  &4.89   \\
 $^{240}$U &5.80& 5.92  &5.59   \\
 $^{238}$Pu &5.30& 5.25  &4.85   \\
 $^{240}$Pu &5.50& 5.92  &4.74   \\
 $^{242}$Pu &5.50& 5.48  &5.25   \\
 $^{244}$Pu &5.30& 5.31  &5.78   \\
 $^{246}$Pu &5.30& 5.06  &6.27   \\
 $^{242}$Cm &5.00& 5.56  &4.24   \\
 $^{244}$Cm &5.00& 5.56  &5.05   \\
 $^{246}$Cm &4.70& 5.40  &5.69   \\
 $^{248}$Cm &5.00& 5.08  &6.07   \\
 $^{250}$Cm &4.40& 4.53  &5.51   \\
\hline
\end{tabular}
\end{table}
We have chosen $^{249}$U, which leads to $^{250}$U with capture
of thermal neutron for the present study. This isotope is only 11
neutrons away from the naturally occuring known Uranium
isotope $^{238}$U,
and therefore likely to be produced with RIB facilities
under construction. The question of how the neutron-rich isotopes
of the actinide nuclei will decay by fission has not been addressed before.
The neutron separation energy
$S_n$ provides a measure of excitation of the compound nucleus.
In our study we use the fission barrier from the extensive
study of Howard and M\"oller (HM) \cite{howard80}
which is being widely used in literature.
It is also the only exhaustive
calculation which gives barrier
for the whole range of known and unknown nuclei where fission decay can
occur. We have compared in Table I
the HM values of
$B_f$ with experiment for 18 actinide nuclei of the four elements
Th, U, Pu and Cm for which data are available \cite{moller95}. These data
are for the
external barriers which in general correspond to the maximum barrier height
calculated by Howard and M\"oller. We have also included in the
table the $B_f$ values calculated by M\"oller et al \cite{moller95}
in finite-range
droplet model (FRDM) for a relative
comparison. It is interesting to see
that HM values are in much  better agreement with experiment than
those of FRDM,
although the latter are comparatively more recent.
For the Uranium isotopes which is of
special interest in the present study, the agreement is remarkably
satisfactory. Therefore we have used them in Fig. 1,
shown as closed circles for all the Uranium isotopes with the
neutron number in the range 140 $\sim$ 180. For the neutron separation energies
$S_n$, we have used the three mass formulae
HM \cite{howard80}, FRDM \cite{moller97} and infinite nuclear matter (INM)
\cite{nayak99}.
It is well known that if $S_n < B_f$, then the nucleus cannot undergo
thermal neutron fission. The fission threshold $E_{nm}=B_f-S_n$ has
to be overcome by impinging with an energetic neutron, in order that
the nucleus will undergo fission.
However, if $S_n > B_f$, then thermal
neutron (with practically zero energy) can cause fission.
It is interesting to see in Fig. 1 that, the isotopes of Uranium in
the range $N=154-172$
are thermally fissile, a feature emanating from the close shell
nature of N=162/164. To show that this feature is not accidental
or specific to Uranium, but a general one arising out of the
close shell structure of N=162 (or 164), we have presented the case of
Th-isotopes in the same figure. 

We now consider the nature of the fission
decay mode of $^{250}$U, which is primarily governed by the
profile of the fission barrier. The height and width of the fission barrier
which is supposed to be parabolic in nature have to be
obtained\cite{vanden73}.
We have followed a semiempirical method to get the width of the barrier
from the systematics of the known experimental fission half-lives,
and extrapolated them to neutron-rich region of interest.
In spontaneous fission the fissioning fragment will see the maximum barrier
which will determine the tunneling probability
more decisively. Hence, the maximum barrier
given by HM \cite{howard80} can be considered as the effective parabolic 
barrier governing the decay, which should be appropriate for the present study.
The fission half-lives can be calculated using the relation \cite{vanden73}
$\tau_{1/2}=ln 2/np$, where n is the number of
barrier assaults by the decaying fragment related to the barrier curvature
energy $\hbar\omega$ by $n\hbar=\hbar\omega/2\pi$, and $p$ is the
penetrability of the barrier given by $p=[1+exp(2\pi B_f/\hbar\omega]^{-1}$.
It may be noted that $\hbar\omega$ is a measure of width of the barrier; 
smaller
the $\hbar\omega$ larger is the barrier width, and hereafter we refer it
as width to bring out a physical picture. Taking the HM values
\cite{howard80} of $B_f$ and using the
experimental $\tau_{1/2}$ \cite{holden00}, we get the values
of $\hbar\omega$. The systematics of $\hbar\omega$ versus
$B_f$ so obtained are shown in Figure 2 for various even-even actinide nuclei.
It is indeed quite
revealing that the plot shows a linear behaviour of $\hbar\omega$
with $B_f$ for a given Z, with
progressively increasing slope with the increase of
proton number of the elements from 90 to 96. For the next element Cf with Z=98,
the linear behavior gets fuzzy. However, the mean follows the trend
with a higher inclination. The trend is more conspicuously restored
for Fm with Z=100. This deviation coincidentally
correlates well with the fission mass yield
systematics, where considerable deviation from the standard
well-defined two peaks occurs for Fm isotopes.
The width $\hbar\omega$ for any isotope with calculated
fission barrier may be
obtained by extrapolation of the linear graph. Since $^{250}$U has
158 neutrons and of considerable distance from the close shell
N=162 (or 164) and itself only 12 neutrons away from the known $^{238}$U
this extrapolation should be quite meaningful and reliable. For $^{250}$U
with a fission barrier of 4.3 MeV \cite{howard80}, we obtained the value of
$\hbar\omega$ as 0.225 MeV from Fig. 2. Now we can construct the
parabolic barrier of base width $\triangle r$ somewhat schematically,
using the value of $\hbar\omega$ following the relation
$1/\hbar\omega=d^2V/dr^2$, where $V$ is the potential energy.
The barrier so obtained is relatively flat and wide compared to the
fission barrier of $^{236}$U with $\hbar\omega=0.357$ MeV, obtained from
the experimental half-life \cite{holden00} and $B_f=5.74$ MeV from Howard and
M\"oller \cite{howard80}.  The normal nuclei like $^{230,232}$Th or
$^{235,238}$U have $\hbar\omega\approx 0.4-0.5$MeV. Thus,
the neutron-rich heavy isotope $^{250}$U has considerably lower value of
$\hbar\omega$, and consequently relatively larger width, arising out of
excess of neutrons. This
flattening of fission barrier makes the nucleus stable against
spontaneous fission decay, because of decreasing penetrability.
The spontaneous fission decay half-life calculated with these
height and width comes out to be $5.7 \times 10^{24}$ years, which
is several order of magnitude higher than that of $^{236}$U.
However, due to the decreased fission barrier of 4.3 MeV for $^{250}$U, with a
minute inducement by a thermal neutron, fission decay will occur.
This exotic feature is due to excess of neutrons, which is not the case in the
normal nuclei in the valley of stability.

Fission studies \cite{samant95,madler85} of $^{235}$U with thermal neutrons
have shown that neutron emission from
the neck region is many times (order of magnitude) larger than the
alpha particle emission. Since clustering probability increases
at low density, alpha particle emission itself is larger compared
to that of proton. Therefore, it is believed that neutron-rich
neck is produced during the scission. This picture is further
reinforced by the polarization effect induced by the Coulomb repulsion
between the two newly formed fragment nuclei. It is also supported
in microscopic studies through our calculation in relativistic mean
field (RMF) theory \cite{arumugam05}
as will be shown aposteriori.
\begin{figure}[ht]
\begin{center}
\includegraphics[width=07cm,height=06cm,angle=0]{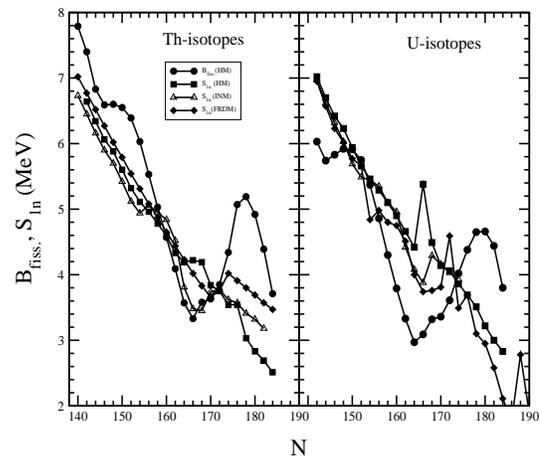}
\end{center}
\caption{Fission barrier $B_f$, and binding energy of the last
neutron $S_{1n}$ as a function of mass number A for Th and U isotopes.
The $B_f$ are taken from \cite{howard80} and $S_{1n}$
are taken from Refs. \cite{howard80},
\cite{nayak99}, \cite{moller95} and \cite{lala99} for HM, INM, FRDM
and RMF model respectively.
}
\label{fig1}
\end{figure}

\begin{figure}[ht]
\begin{center}
\includegraphics[width=7cm,height=6cm,angle=0]{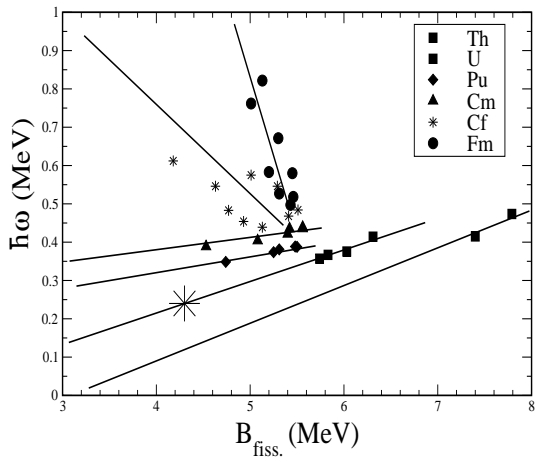}
\end{center}
\caption{The barrier curvature energy $\hbar\omega$ as function of 
fission barrier $B_f$.  The $B_f$ values and the experimental half-lives 
taken from Howard and M\"oller \cite{howard80} and  Ref. \cite{holden00}, 
respectively.
The star on the line for Uranium in figure,
denotes the value of $\hbar\omega$ to be 0.225 MeV for its $B_f=4.3$ MeV
corresponding to $^{250}$U.}
\label{fig2}
\end{figure}
\begin{figure}[ht]
\begin{center}
\includegraphics[width=7cm,height=6cm,angle=0]{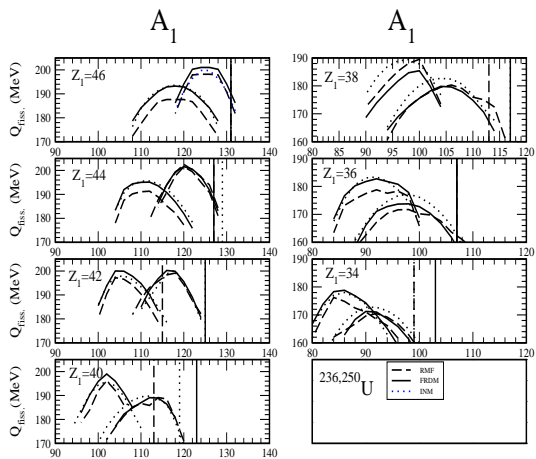}
\end{center}
\caption{$Q_{fiss.}-$value distribution 
for $^{236}$U and $^{250}$U as a function of $A_1$
fragment in the binary decay $A\rightarrow A_1+A_2$.
The binding energy used for calculation of $Q-$value is
taken from \cite{lala99}, \cite{moller97} and Refs. \cite{nayak99}
for RMF, FRDM and INM, respectively. 
The vertical line marks the
neutron drip-line for the corresponding element in each panel.
}
\label{fig3}
\end{figure}
\begin{figure}[ht]
\begin{center}
\includegraphics[width=8cm,height=6cm,angle=0]{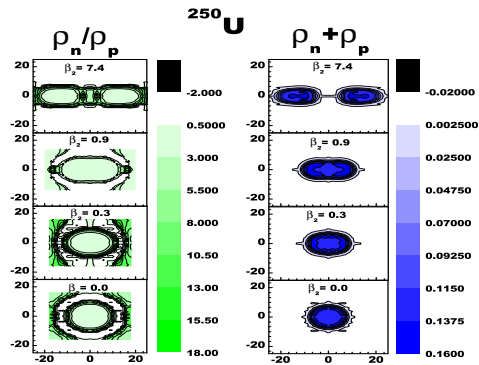}
\end{center}
\caption{The matter density distributions $\rho=\rho_n+\rho_p$ (in fm$^{-3}$)
and $\alpha=\rho_n/\rho_p$ of $^{250}$U obtained
in the RMF calculations with NL3 force.}
\label{fig4}
\end{figure}
As is well known, the major driving force for the decay of a nucleus is the
$Q-$value of the reaction. The probability of fragment mass yield
in a given channel is directly related to the $Q-$value. We, therefore,
calculate here the $Q-$value systematics of the fission of a nucleus
(A,Z) decaying to ($A_1,Z_1$) and ($A_2,Z_2$) defined as
$Q^f(A,Z)=BE(A_1,Z_1)+BE(A_2,Z_2)-BE(A,Z)$.
In Figure 3, we have plotted the $Q-$values of the
binary decay into two fragments $A_1$ and $A_2$, as a function of the
mass number of the $A_1$ fragment,
for all the relevant elements with even values of $Z_1$, starting
from 34 to 46.  The complimentary fragment ($Z_2$, $A_2$) is thereby fixed.
Since the yield falls rapidly
with the decrease in $Q-$value for an element, we have only shown the
distribution of $Q-$values lying
above $90\%$ of the highest values. For the sake of comparison, the
$Q-$value distributions for both $^{236}$U and $^{250}$U are
shown in Figure 3. To ensure that our conclusions remain quite
general and valid, and independent of any specific mass formula, we have
used the masses predicted in the three mass models RMF \cite{lala99}, 
FRDM and INM.  The Q-value distributions for $^{236,250}$U are presented 
in Figure 3 as dashed, solid and dotted curves for RMF, FRDM and INM 
mass formulae respectively.
In the figure, three corresponding vertical lines mark the
drip-lines of the respective elements predicted in the three mass models.
It can be seen that the drip lines for all the three mass models agree within
$\sim 3-5$ neutrons, except in case of Z=40 isotopic chain for the
FRDM mass model. In most of the cases, the drip lines fall inside
the $Q-$value distributions and in some cases they touch the outer fringe
shown in the figure with the exception of Z=34 and 40 where the FRDM drip
lines are somewhat away.  Thus
all the isotopes lying to the right of the drip lines will be
unstable against spontaneous release of neutrons from the
fragments at scission which is not the case with $^{236}$U for which
the drip lines are far away (See Fig. 3).
In the usual fission process of $^{236}$U, neutrons are emitted
from the fragments after they are fully accelerated. But in the
present case of $^{250}$U, a certain number of neutrons will be simultaneously
produced along with the two heavy fragments signaling a new mode of
fission decay which may be
termed as {\it multi-fragmentation fission}. An order of magnitude of these
prompt multi-fragmentation neutrons as estimated from the mass yield plot for
$^{250}$U shown in Figure 3, turns out to be about
2 to 3 neutrons per fission. These are the additional neutrons apart
from the normal multiplicity of neutrons emitted from
the fragments. In case of neutron-induced fission of normal
$^{233}$U and $^{235}$U nuclei, the fission neutron multiplicities are
of the order $2.5$ \cite{samant95}.
This number is, therefore, more than doubled in case of
$^{250}$U fission, which will have important implications on the
energetics of the fission process. This phenomenon will be more prominent
in heavier Uranium isotopes due to availability of more neutrons.
Although, the beta decay life-time of $^{250}$U is few tens of seconds
\cite{moller97} it is much larger than the nucleon decay life time, which is
of the order of $10^{-17}$ seconds and therefore it will have
an implication in the $r-$process nuclear synthesis
and consequently stellar evolution.

We used RMF theory with NL3 interaction to study the
evoluation of density as the nucleus undergoes distorsion in
its shape on its path to fission, to see if microscopic calculation
supports the above picture. Recently, calculations on nuclear
densities obtained in the RMF studies have yielded useful insight and
results on nuclear structure and dynamics \cite{arumugam05}, which
we follow here. With this 
view we carried out RMF calculations for successively increasing 
deformation $\beta_2$ starting from the ground state.
Now we present our results of such calculations on the
total (neutrons + protons) matter density distributions
$\rho=\rho_n+\rho_p$, and the ratio of neutron to proton
density variable $\alpha=\rho_n/\rho_p$
as function of the deformation parameter $\beta_2$.
In Figures 4, we have presented the results of such calculations
for $^{250}$U, with specified column
on the right of the distributions showing the scale through
different colours.
From the figure it is clear that $^{250}$U gets more and more
elongated as in the usual liquid drop picture of fission
and finally splits into two
parts with a neck connecting them for the
deformation $\beta_2=7.4$. It is interesting to examine the
composition of the neck in this configuration. The density of the
neck obtained in our
calculation and also evident from the picture is $\rho=0.02383 fm^{-3}$
which is quite low as expected. And the neutron to proton
density ratio is obtained as $\rho_n/\rho_p=3.54703$ which
can be contrasted with the average density ratio of $\sim 1.54$ in 
its ground state. Thus these microscopic studies
coroborate the neutron rich-ness of neck as found in experiment
and other studies \cite{samant95,madler85}. Similar studies
for $^{236}$U gives the
density of the neck $\rho=0.02440 fm^{-3}$ and
the neutron-proton density ratio $\rho_n/\rho_p= 2.73089$.
In case of $^{250}$U, the neck is much richer in neutron
since 14 extra neutrons are available
compared to $^{236}$U which is in conformity with expectation.
This favours simultaneous emission of
neutrons along with the two fragments at scission, strongly
supporting the multifragmentation fission process predicted above.

In conclusion, we have identified the chain of Uranium isotopes
with neutron numbers N=154 to 172 which are thermally fissile.
This is a likely manifestation of the close shell nature of the
magic number 162/164.  We have chosen $^{249}$U 
as a representative nucleus to study the thermal fission decay.
The fission decay properties of such neutron-rich nuclei, in particular
in the actinide region away from the valley of stability, have not been
addressed before, which has become important in the context of
RIB programs. Its fission barrier 
profile has been shown to be relatively flat and wide compared to $^{236}$U
yielding a half-life of $5.7\times 10^{24}$ years, which makes it
extremely stable against spontaneous fission, while rendering it highly
vulnerable to thermal neutron fission$-$ a unique property indeed.
On the basis of the probability of mass yield and microscopic RMF
calculations, strong evidence of a new mode of fission decay is revealed, where
in addition to two heavy fragments,
2 to 3 scission neutrons will be simultaneously emitted, which may be
termed as {\it {multifragmentation fission}}. 
These extra neutrons are in addition to the normal multiplicity of
neutrons emitted by the
excited fission fragments which is of the order of 2.3 to 2.5.
Thus the  doubling of the
neutron emission per fission will have implications for the $r-$process
nucleosynthesis in the steller evolution. Whether such a nucleus presents an
attractive possibility as a source of energy production needs to be
examined. The above phenomenon is a general one, not restricted to $^{250}$U
and is likely to be more prominent in heavier Uranium
isotopes.


\begin{thebibliography}{99}
\bibitem{thoen04} M. Thoennessen, Rep. Prog. Phys. {\bf 67}, 1187 (2004).
\bibitem{howard80} W.M. Howard and P. M\"oller, At. Data and Nucl. 
                 Data Tables, {\bf 25}, 219 (1980).
\bibitem{cwiok96} 
           G. M\"unzenberg and S. Hofmann, {\it Heavy elements and related 
	   new phenomena (World Scientific, 1999)} and references there in, 
	   Eds: W. Greiner and R.K. Gupta,
	   Ch. 1, page 9.
\bibitem{moller97}	   
           P. M\"oller, R.J. Nix and K.-L. Kratz,
           At. Data and Nucl. Data Tables, {\bf 66}, 131 (1997).
\bibitem{moller95}
           P. M\"oller, R.J. Nix, W.D. Myers and W.J. Swiatecki,
           At. Data and Nucl. Data Tables, {\bf 59}, 185 (1995).
\bibitem{nayak99} R.C. Nayak and L. Satpathy,
        At. Data and Nucl. Data Tables, {\bf 73}, 213 (1999).
\bibitem{vanden73} R. Vandenbosch and J.R. Huizenga,
           {\it Nuclear Fission}, Academic press, inc. (1973) Ch. III,
           p. 45.
\bibitem{holden00} N.E. Holden and D.C. Hoffman,
           Pure Appl. Chem. {\bf 72}, 1525 (2000).
\bibitem{samant95} M.S. Samant et al., Phys. Rev. {\bf C51}, 3127 (1995).
\bibitem{madler85} P. Madler, Z. Phys. {\bf A321}, 343 (1985).
\bibitem{lala99} G.A. Lalazissis, S. Raman and P. Ring,
        At. Data and Nucl. Data Tables, {\bf 71}, 1 (1999).
\bibitem{arumugam05} P. Arumugam et al., Phys. Rev. {\bf C71}, 064308 (2005);
        B.K. Sharma, P. Arumugam, S.K. Patra, P.D. Stevenson,
        Raj K. Gupta and W. Greiner,
        J. Phys. {\bf G32}, L1 (2006); Raj K. Gupta, S.K. Patra,
        P.D. Stevenson and W. Greiner, Int. J. Mod. Phys. {\bf E} (in press).



\end{thebibliography}
\end{document}